# A deep convolutional encoder-decoder neural network in assisting seismic horizon tracking

*Hao Wu\*, Bo Zhang, University of Alabama*

**Summary**

Seismic horizons are geologically significant surfaces that can be used for building geology structure and stratigraphy models. However, horizon tracking in 3D seismic data is a time-consuming and challenging problem. Relief human from the tedious seismic interpretation is one of the hot research topics. We proposed a novel automatically seismic horizon tracking method by using a deep convolutional neural network. We employ a state-of-art end-to-end semantic segmentation method to track the seismic horizons automatically. Experiment result shows that our proposed neural network can automatically track multiple horizons simultaneously. We validate the effectiveness and robustness of our proposed method by comparing automatically tracked horizons with manually picked horizons.

**Introduction**

Seismic horizon tracking is an important step for seismic interpretation. The seismic horizons can be treated as the stratigraphic boundaries which can represent the depositional environments and geological features. Generally, we track the horizons by picking the similar configuration of samples such as peaks, troughs, or zero-crossing points through the consistently seismic traces. Manually tracking is the most familiar but the least efficient interpretation techniques.

In recent decades, many approaches have been proposed to build the horizon model without manual picking. These automatically methods can be classified into three categories (Wu and Hale, 2015). Zeng (1998) proposed a method which first manually picking several reference horizons slice and then interpolated a horizon volume. Lomask (2006) first calculated the local dips over the entire seismic volume and transferred them into the time shift and then apply least-square method to track the horizons automatically. Stark (2003) used the instantaneous unwrapping phase to generate a relative geological time volume and applied to automatically seismic horizon tracking. However, those methods assume that the horizon is locally smooth between two control points. Horizons are inaccurate near fault or other complicated geological zones.

Deep learning has attracted significant attention in geoscience field. Convolution neural network is one of the most popular and widely used deep learning algorithms. CNNs (LeCun, 1989) effectively learn different scale image features in the images. Long et al. (2014) defined the current general encoder-decoder semantic segmentation neural network architecture. The encoder-decoder architecture is a breakthrough in the history of deeply learning algorithm. The encoder is a pre-trained classification network and decoder is to project the discriminative features learned by the encoder semantically. Badrinarayanan (2015) presented SegNet algorithm for pixel-wise semantic segmentation. The Segnet has two advantages. The first advantages is using upsampling layer in the decoder to keep high-frequency detail intact in the segmentation. The second advantage is using convolutional layers instead of using fully connected layer, which can memory the indices of image features.

The process of seismic horizon interpretation can be viewed as the segmentation of seismic traces into different piece wise segments. In this paper, we employ CNN to segment the seismic traces into different zones automatically. The boundaries between different zones are the seismic horizons. We first manually pick several horizons to build the ground truth horizons. We then randomly choose 1% seismic traces as the train data and the rest 99% seismic traces as the test data.

**Method**

We randomly choose 1% seismic traces within the seismic survey as the training data. The interpreted horizons divide the seismic traces into different segment. We give the same labels to the segment of seismic traces which are bounded the same horizons. The rest seismic traces function as the testing data set used for validating the effectiveness of our algorithm. The curve patterns of the seismic amplitude can be treated as the image features. Figure 1 shows the automatically seismic horizon interpretation workflow using the deep convolutional encoder-decoder neural network. The proposed network consists of two main parts: encoder and decoder networks. The role of encoder network is learning the features imbedded in each segment of seismic traces of the training set. The decoder network is automatically segment the seismic traces using the learned features in the decoder process. The decoder process consists four layers and each layer contains convolution layer, batch-normalization layer, activation layer and max-pooling layer. The decoder process also consists four layers which corresponded to the encoder network. Each layer in the decoder contains up sampling layer, convolution layer, batch-normalization layer and activation layer. The pixel-wise classification layer is the last process in our workflow. The output of our workflow is segmented seismic traces and the boundary of each segments are the automatically tracked horizons.

# Deep learning in assisting seismic horizon tracking

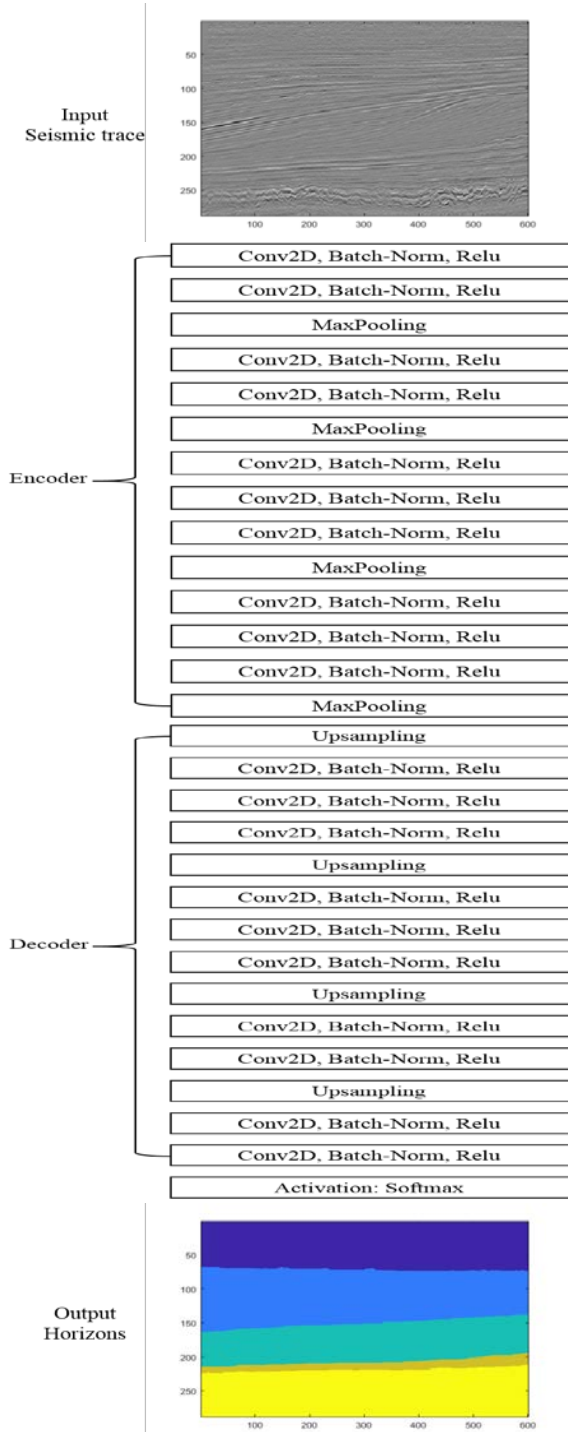

Figure 1. The proposed workflow for automatically horizon picking by using deep convolutional neural network.

### Application

We demonstrate the effectiveness of our workflow using the F3 block seismic data acquired over the offshore North Sea, Netherland. The seismic survey consists of 651 inline and 951 crossline. The time sample increment is 4ms. The project provided by Opendtect already have four interpreted horizons (Figure 2).

### *Data preparation for encoder-decoder training*

To train a deep convolutional encoder-decoder neural network, we need to organize our dataset and feed into the network. We first need label the seismic tracesbegan with manually picking four horizons to build a horizon model (Figure 2). We then labeled the seismic traces used for training according to interpreted horizons. There are four horizons in our case and the seismic traces are segmented into four segments. There are 5 labels corresponding to the segmented seismic traces. Figures 3a and 3b are the seismic inline sections before and after labeling according to the interpreted horizons. The training seismic traces account 1% for all the seismic traces and we randomly select them from the whole seismic survey. The rest of 99% seismic traces were treated as test seismic traces used for demonstrating the accuracy of our proposed workflow.

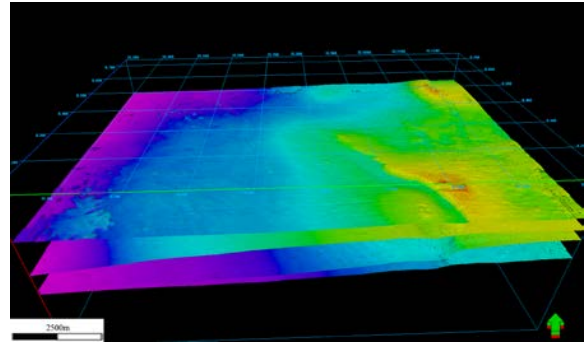

Figure 2. Manually picked horizon volume.

# Deep learning in assisting seismic horizon tracking

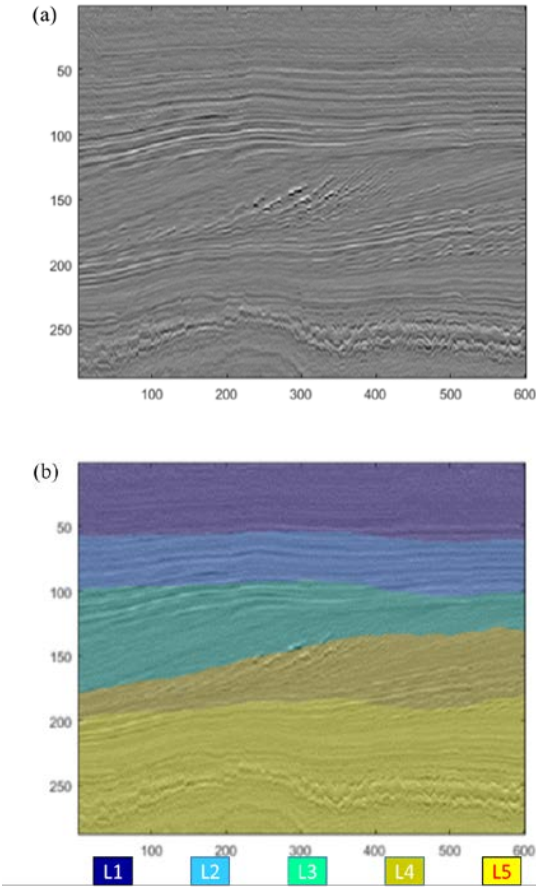

Figure 3. A section of seismic traces and labeled traces. (a) The seismic traces. (b) The seismic traces and their corresponding labels.

### Build an encoder-decoder neural network

We next built a deep convolutional encoder-decoder neural network. There are four layers in the encoder network and size of convolution kernels are 32, 16, 8, and 3, respectively. In the decoder network, we also built four layers that correspond to the encoder network. The size of the convolution kernels in the decoder network are 3, 8, 16, and 32, respectively. The training seismic traces and corresponding segment labels are the inputs for the encoder-decoder neural network training. Our training process converges after 10 epoches. The validation accuracy in the training process is above 99%.

### Evaluate the prediction

To evaluate the result, we compared the predicted horizons with the ground truth horizons. We first use our trained neural network to segment the test seismic traces. We then extract the horizons according to the boundary between two segments. Figure 4 show the extracted four horizons in 3D view. Figure 5 shows predicted and "ground truth" horizons overlaid on a representative inline section. The red and blue curves are the predicted and "ground truth" horizons, respectively. Note that perfect match between red and blue horizons. The average absolute error between the ground truth horizons and predicted horizons is 1.13ms. Figure 6 compares interpolated and predicted horizons. The blue, red, and green curves in Figure 6 are the "ground truth", predicted, and interpolated horizons, respectively. Note that our predict horizon successfully follows the seismic reflections across the faults. However, the interpolated horizon fails to follow the local reflectors.

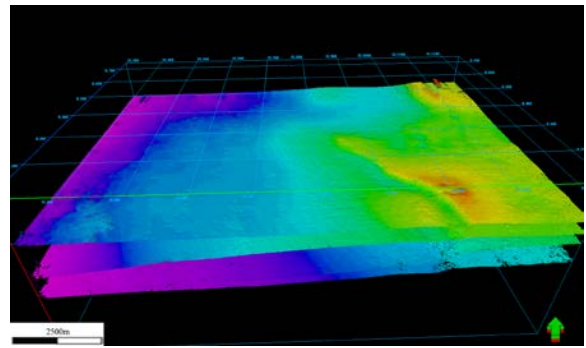

Figure 4. Predicted horizon volume.

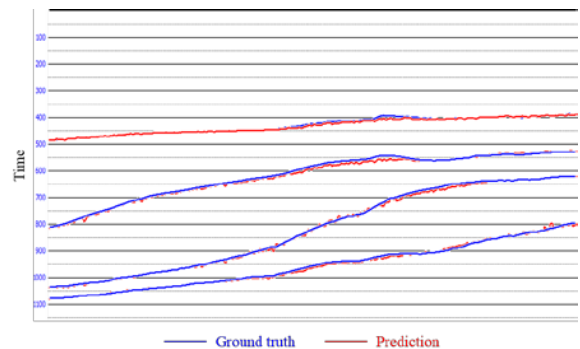

Figure 5. Comparison between the ground truth horizons and predicted horizons.

# Deep learning in assisting seismic horizon tracking

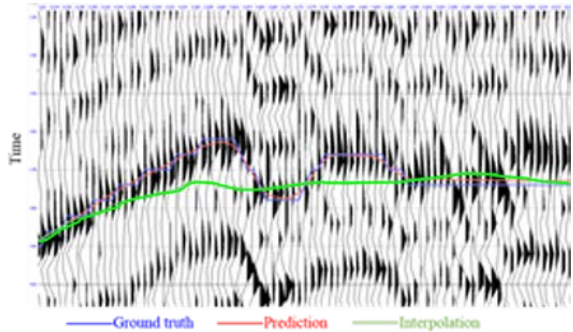

Figure 6. Comparison between the interpolated horizon and predicted horizon. Note that the predicted horizons have a better performance at complicated structure.

**Conclusion**

We have presented a novel semi-automatically seismic horizon tracking method by using a state-of-art end-to-end semantic segmentation technique. We first randomly choose some seismic traces to form the training data set. We segment the training seismic traces into different zone according to the interpreted horizons and give a label to each of zone accordingly. We then build a deep convolutional encoder-decoder neural network. We next input the train data and corresponding label into the network for training. We finally apply the trained network to the test data to segment the seismic traces. The result shows that the average absolute error between our predicted horizons and the ground truth horizons is smaller than the value of one time sample. The comparison between our predicted horizons and the interpolated horizons illustrates that our proposed method has a better performance at the complicated geological structure zone.

**Acknowledgments**

The authors thank the CGG providing the academic license for Hampson-Russell software. The authors also thank the technical discussion with Dr. Tao Zhao from Geophysical insights.